# Generalised Rabin(1) synthesis[*][†]

Rüdiger Ehlers

Saarland University

October 29, 2018

We present a novel method for the synthesis of finite state systems that is a generalisation of the generalised reactivity(1) synthesis approach by Piterman et al. (2006). In particular, we describe an efficient method to synthesize systems from linear-time temporal logic specifications of the form

$$(a_1 \wedge a_2 \wedge \ldots \wedge a_{n_a}) \rightarrow (g_1 \wedge g_2 \wedge \ldots \wedge g_{n_g})$$

for which each of the assumptions $a_i$ and guarantees $g_i$ has a Rabin index of one. We show how to build a parity game with at most five colours that captures all solutions to the synthesis problem from such a specification. This parity game has a structure that is amenable to symbolic implementations. We furthermore show that the results obtained are in some sense tight, i.e., that there does not exist a similar synthesis method for assumptions and specifications of higher Rabin index, provided that $P \neq NP$.

## 1 Introduction

Synthesis of finite state systems (Kupferman and Vardi, 1999) has been proven to be a valuable concept for the development of open systems that are correct by construction. In contrast to verification, it frees the designer of a computation system from the task to actually build the system in addition to stating its specification. Therefore, this technique can significantly reduce the time for developing a correct system, making it attractive in practice.

The first works in this area were concerned with *closed synthesis*, where everything that can be be reasoned about is under the control of the system to be synthesized. More recent works are concerned with *open synthesis*. Here, there exists some input to the system which is not under its control. This model is more suitable for synthesis of *reactive systems*, as almost all such systems of practical relevance have some uncontrollable input. In this context, *linear-time temporal logic* (LTL, see, e.g., Vardi, 1996) is the predominant specification language used.

One drawback of synthesis is that its time complexity for LTL specifications is doubly-exponential in the length of the specification (Pnueli and Rosner, 1989), making the problem intractable in general. One of the reasons for this high complexity is the fact that it is possible to formulate specifications for which the smallest implementation satisfying it is of size doubly-exponential in the length of the specification. More recently, it has been argued that such high size bounds rarely occur for specifications used in practice (Jobstmann and Bloem, 2006; Schewe and Finkbeiner, 2007), so this does not necessarily affect the efficiency of synthesis for practical applications.

---

[*]This work was supported by the German Research Foundation (DFG) within the program "Performance Guarantees for Computer Systems" and the Transregional Collaborative Research Center "Automatic Verification and Analysis of Complex Systems" (SFB/TR 14 AVACS).

[†]This is the second revision of this paper. In comparison to the first version, the main construction in Section 4 has been simplified and some changes in style of writing have been performed.

Apart from approaches for synthesis from arbitrary LTL formulas, there also exist specialised procedures for specifications of certain forms. In particular, it has been observed that many specifications found in practice are of the form $\psi \rightarrow \phi$ for some conjunctions of safety and basic liveness properties $\psi$ and $\phi$. We call $\psi$ the assumption part of such a specification and $\phi$ the guarantee part. A basic liveness property is a conjunct that can be represented by the LTL formula $GFp$ for some atomic proposition $p$. Piterman et al. (2006) were able to show that the synthesis problem for such *generalised reactivity(1)* formulas can be solved in time cubic in the state space of the design. Subsequent works (Bloem et al., 2007b,a) showed that indeed this approach can be used successfully in practice.

Recently, it has been observed (Sohail et al., 2008) that the low complexity of the synthesis problem for generalised reactivity(1) specifications is not surprising as the problem can be reduced to solving a parity game with precisely 3 colours. Furthermore, the state space of this parity game is (almost) the product state space of the deterministic Büchi automata representing the individual conjuncts of the assumption and guarantee parts, making this approach amenable to the symbolic solution of the game, for example using the algorithm by McNaughton/Zielonka (see Grädel et al., 2002 or Schewe, 2008 for comprehensive descriptions) with *binary decision diagrams* (see, e.g., Baier and Katoen, 2008).

As the set of properties representable by generalised reactivity(1) formulas is still relatively limited (for example, it cannot be specified that the system to be constructed should have some finite initialisation period after which it must output "ready" forever), a natural question to ask is if this approach can be extended in order to include more representable properties without losing the possibility to encode the overall specification of the system as a parity automaton with a constant number of colours. More precisely, we ask the question what type the assumption and guarantee conjuncts may be of such that we can build a deterministic parity automaton of size polynomial in the product of the automata representing the individual assumptions and guarantees and its induced parity game is won if and only if the overall specification is realisable and the number of colours is constant (and independent of the number of assumption and specification conjuncts).

In this paper, we present an answer to this question. In fact, the constant number of colours can be retained if the assumption and guarantee conjuncts have a Rabin index of 1, leading to five colours in total. We call this approach *generalised Rabin(1) synthesis*. This result is strict, i.e., for every Rabin index $\geq 2$, a constant number of colours does not suffice (for only a polynomial blow-up in the state space of the automaton), unless $P = NP$. The added expressivity is shown to be of value for practical cases, making this approach a practically suitable trade-off between the approaches allowing full LTL for the specification and the faster frameworks.

We start by giving the basic definitions in Section 2. Then, we discuss how to convert the individual conjuncts in the specification to Rabin automata with a Rabin index of 1. Section 4 discusses how a parity game that captures the synthesis problem for such specifications can be built. Section 5 then discusses the possibility for similar approaches to specifications with conjuncts of higher Rabin index and contains the corresponding negative result. Section 6 sketches an application domain that benefits from the extended applicability of generalised Rabin(1) synthesis in comparison to generalised reactivity(1) synthesis. Section 7 finally concludes and gives an outlook.

## 2 Preliminaries

**Words, Languages and natural numbers**  Let $\Sigma$ be a finite set. By $\Sigma^*/\Sigma^\omega$ we denote the set of all finite/infinite sequences, respectively. Such sequences are also called *words* over $\Sigma$. Sets of words are also called *languages*. For the scope of this paper, we denote the set of natural numbers including 0 by $\mathbb{N}_0$. For simplicity, if 0 is excluded, we simply write $\mathbb{N}$.

For some sequence $w = w_0 w_1 \ldots$, we denote by $w^j$ the suffix of $w$ starting with the $j$th symbol, i.e., $w^j = w_j w_{j+1} \ldots$ for all $j \in \mathbb{N}_0$.

**Mealy automata**  *Reactive systems* are usually described using a *finite state machine* description. Formally, we define *Mealy automata* as five-tuples $\mathcal{M} = (S, \Sigma_I, \Sigma_O, \delta, s_0)$ where $S$ is some finite set of states, $\Sigma_I$ and $\Sigma_O$ are input/output alphabets, respectively, $s_0 \in S$ is the initial state and $\delta : S \times \Sigma_I \to S \times \Sigma_O$ is the transition function of $\mathcal{M}$. The computation steps of a Mealy automaton are called *cycles*.

For the scope of this paper, we usually set $\Sigma_I = 2^{AP_I}$ and $\Sigma_O = 2^{AP_O}$ for some sets of input/output atomic propositions $AP_I$ and $AP_O$. This is a typical choice in literature on synthesis and verification (Kupferman and Vardi, 1999, 1997; Vardi, 1996; Schewe and Finkbeiner, 2007; Bloem et al., 2009; Filiot et al., 2009) as specification logics such as LTL are usually used to describe behaviour of the system with respect to the individual atomic propositions and Mealy automata implemented in hardware usually have such an input/output structure (in which the individual atomic propositions represent the values of the input/output *signals* of the system).

**The language induced by Mealy automata**  Given a Mealy automaton $\mathcal{M} = (S, \Sigma_I, \Sigma_O, \delta, s_0)$ and some input word $i = i_0 i_1 \ldots \in \Sigma_I^\omega$, $\mathcal{M}$ induces a *run* $\pi = \pi_0 \pi_1 \ldots$ and some *output word* $o = o_0 o_1 \ldots$ over $i$ such that $\pi_0 = s_0$ and for all $j \in \mathbb{N}_0$: $\delta(\pi_j, i_j) = (\pi_{j+1}, o_j)$. Formally, we define the language of $\mathcal{M}$, written as $\mathcal{L}(\mathcal{M})$, to be the set of words $w = w_0 w_1 \ldots \in \Sigma^\omega$ with $\Sigma = 2^{AP_I \uplus AP_O}$ such that $\mathcal{M}$ induces a run $\pi$ over the input word $i = w|_{\Sigma_I} = (w_0 \cap \Sigma_I)(w_1 \cap \Sigma_I) \ldots$ such that $w|_{\Sigma_O} = (w_0 \cap \Sigma_O)(w_1 \cap \Sigma_O) \ldots$ is the output word corresponding to $\pi$.

**Linear-time temporal logic**  Before a system that is correct with respect to its specification can be synthesized, the specification has to be formally stated. For such a task, *linear-time temporal logic* (LTL) is a commonly used logic. Syntactically, LTL formulas are defined inductively as follows (over some set of atomic propositions $AP$):

- For all atomic propositions $x \in AP$, $x$ is an LTL formula.
- Let $\phi_1$ and $\phi_2$ be LTL formulas. Then $\neg \phi_1$, $(\phi_1 \vee \phi_2)$, $(\phi_1 \wedge \phi_2)$, $X\phi_1$, $F\phi_1$, $G\phi_1$, and $(\phi_1 U \phi_2)$ are also valid LTL formula.

The validity of an LTL formula $\phi$ over $AP$ is defined inductively with respect to an infinite trace $w = w_0 w_1 \ldots \in (2^{AP})^\omega$. Let $\phi_1$ and $\phi_2$ be LTL formulas. We set:

- $w \models p$ if and only if (iff) $p \in w_0$ for $p \in AP$
- $w \models \neg \psi$ iff not $w \models \psi$
- $w \models (\phi_1 \vee \phi_2)$ iff $w \models \phi_1$ or $w \models \phi_2$
- $w \models (\phi_1 \wedge \phi_2)$ iff $w \models \phi_1$ and $w \models \phi_2$
- $w \models X\phi_1$ iff $w^1 \models \phi_1$

- $w \models G\phi_1$ iff for all $i \in \mathbb{N}_0$, $w^i \models \phi_1$
- $w \models F\phi_1$ iff there exists some $i \in \mathbb{N}_0$ such that $w^i \models \phi_1$
- $w \models (\phi_1 U \phi_2)$ iff there exists some $i \in \mathbb{N}_0$ such that for all $0 \leq j < i$, $w^j \models \phi_1$ and $w^i \models \phi_2$

We use the usual precedence rules for LTL formulas in order to be able to omit unnecessary braces and also allow the abbreviations typically used for Boolean logic, e.g., that $a \to b$ is equivalent to $\neg a \vee b$ for all formulas $a$, $b$.

As an example, consider the specification $\phi = G(request \to grant)$ over $AP = \{request, grant\}$. Intuitively, such a specification would be satisfied by all runs of a Mealy automaton $\mathcal{M} = (S, \Sigma_I, \Sigma_O, \delta, s_0)$ with $\Sigma_I = 2^{\{request\}}$ and $\Sigma_O = 2^{\{grant\}}$ if all requests given to $\mathcal{M}$ are answered by a grant immediately. In other papers (e.g., Filiot et al., 2009; Jobstmann and Bloem, 2006), in which the order of input and output is inverted, the specification would have to be changed to $\phi = G(request \to X grant)$ in order to be semantically equivalent to our model here. We however prefer our model as it typically shortens the LTL formulas to be considered in the synthesis procedure.

**Labelled parity games** A labelled parity game is a tuple $\mathcal{G} = (V_0, V_1, \Sigma_0, \Sigma_1, E_0, E_1, v_0, c)$ with $E_0 : V_0 \times \Sigma_0 \to V_1$ and $E_1 : V_1 \times \Sigma_1 \to V_0$. We abbreviate $V = V_0 \uplus V_1$. We only consider finite games here, for which $V_0$, $V_1$, $\Sigma_0$ and $\Sigma_1$ are finite. The initial vertex $v_0$ is always a member of $V_0$. The colouring function $c : V_0 \to \mathbb{N}_0$ assigns to each vertex in $V_0$ a colour. For the scope of this paper, we only assign colours to vertices of player 0.

A decision sequence in $\mathcal{G}$ is a sequence $\rho = \rho_0^0 \rho_0^1 \rho_1^0 \rho_1^1 \ldots$ such that for all $i \in \mathbb{N}_0$, $\rho_i^0 \in \Sigma_0$ and $\rho_i^1 \in \Sigma_1$. A decision sequence $\rho$ induces an infinite play $\pi = \pi_0^0 \pi_0^1 \pi_1^0 \pi_1^1 \ldots$ if $\pi_0^0 = v_0$ and for all $i \in \mathbb{N}_0$, $p \in \{0, 1\}$, $E_p(\pi_i^p, \rho_i^p) = \pi_{i+p}^{1-p}$.

Given a play $\pi = \pi_0^0 \pi_0^1 \pi_1^0 \pi_1^1 \ldots$, we say that $\pi$ is winning for player 0 if $\max\{c(v) \mid v \in V_0, v \in \inf(\pi_0^0 \pi_1^0 \ldots)\}$ is even for the function $\inf$ mapping a sequence onto the set of elements that appear infinitely often in the sequence. If a play is not winning for player 0, it is winning for player 1.

Given some parity game $\mathcal{G} = (V_0, V_1, \Sigma_0, \Sigma_1, E_0, E_1, v_0, \mathcal{F})$, a strategy for player 0 is a function $f : (\Sigma_0 \times \Sigma_1)^* \to \Sigma_0$. Likewise, a strategy for player 1 is a function $f : (\Sigma_0 \times \Sigma_1)^* \times \Sigma_0 \to \Sigma_1$. In both cases, a strategy maps prefix decision sequences to an action to be chosen next. A decision sequence $\rho = \rho_0^0 \rho_0^1 \rho_1^0 \rho_1^1 \ldots$ is said to be in correspondence with $f$ if for every $i \in \mathbb{N}_0$, we have $\rho_n^p = f(\rho_0^0 \rho_0^1 \ldots \rho_{n+p-1}^{1-p})$. A strategy is winning for player $p$ if all plays in the game that are induced by some decision sequence that is in correspondence to $f$ are winning for player $p$. It is a well-known fact that for parity games, there exists a winning strategy for precisely one of the players (see, e.g., Grädel et al., 2002). We call a state $v \in V_0$ winning for player $p$ if changing the initial state to $v$ makes or leaves the game winning for player $p$. Likewise, a state $v' \in V_1$ is called winning for player $p$ if a modified version of the game, that results from introducing a new initial state with only one transition to $v'$ is (still) winning for player $p$.

If a strategy $f$ for player $p$ is a *positional strategy*, then $f(\rho_0^0 \rho_0^1 \ldots \rho_n^p) = f'(E_{1-p}(\ldots E_1(E_0(v_0, \rho_0^0), \rho_0^1), \ldots, \rho_{n+p-1}^{1-p}))$ for some function $f' : V_p \to \Sigma_p$. By abuse of notation, we call both $f'$ and $f$ positional strategies. Note that such a function $f'$ is finitely representable as both domain and co-domain are finite. For parity games, it is known then there exists a winning positional strategy for a player if and only if there exists some winning strategy for the same player.

Note that a translation between this model and an alternative model where the colouring function is defined for both players is easily possible with only a slight alteration of the game structure.

**$\omega$-automata** An $\omega$-automaton $\mathcal{A} = (Q, \Sigma, q_0, \delta, \mathcal{F})$ is a five-tuple consisting some finite state set $Q$, some finite alphabet $\Sigma$, some initial state $q_0 \in Q$, some transition function $\delta : Q \times \Sigma \to 2^Q$ and some *acceptance component* $\mathcal{F}$ (to be defined later). We say that an automaton is deterministic if for every $q \in Q$ and $x \in \Sigma$, $|\delta(q,x)| \leq 1$. Given an $\omega$-automaton $\mathcal{A} = (Q, \Sigma, q_0, \delta, \mathcal{F})$, we also call $(Q, \Sigma, q_0, \delta)$ the *transition structure* of $\mathcal{A}$.

Given an infinite word $w = w_1 w_2 \ldots \in \Sigma^\omega$ and an $\omega$-automaton $\mathcal{A} = (Q, \Sigma, q_0, \delta, \mathcal{F})$, we say that some sequence $\pi = \pi_0 \pi_1 \ldots$ is a run for $w$ if $\pi_0 = q_0$ and for all $i \in \{1, 2, \ldots\}$, $\pi_i \in \delta(\pi_{i-1}, w_i)$. We say that $\pi$ is accepting if for $\inf(\pi) = \{q \in Q \mid \exists^\infty j \in \mathbb{N} : \pi_j = q\}$, $\inf(\pi)$ is accepted by $\mathcal{F}$. The acceptance of $\pi$ by $\mathcal{F}$ is defined with respect to the type of $\mathcal{F}$, for which many have been proposed in the literature (Grädel et al., 2002).

- For a *safety winning condition*, all infinite runs are accepting. In this case, the $\mathcal{F}$-symbol can also be omitted from the automaton definition.

- For a *Büchi acceptance condition* $\mathcal{F} \subseteq Q$, $\pi$ is accepting if $\inf(\pi) \cap \mathcal{F} \neq \emptyset$. Here, $\mathcal{F}$ is also called the set of *accepting states*.

- For a *co-Büchi acceptance condition* $\mathcal{F} \subseteq Q$, $\pi$ is accepting if $\inf(\pi) \cap \mathcal{F} = \emptyset$. Here, $\mathcal{F}$ is also called the set of *rejecting states*.

- For a *generalised Büchi acceptance condition* $\mathcal{F} \subseteq 2^Q$, $\pi$ is accepting if for all $F \in \mathcal{F}$, $\inf(\pi) \cap F \neq \emptyset$.

- For a *parity acceptance condition*, $\mathcal{F} : Q \to \mathbb{N}_0$ and $\pi$ is accepting in the case that $\max\{\mathcal{F}(v) \mid v \in \inf(\pi)\}$ is even.

- For a *Rabin acceptance condition* $\mathcal{F} \subseteq 2^Q \times 2^Q$, $\pi$ is accepting if for $\mathcal{F} = \{(F_1, G_1), \ldots, (F_n, G_n)\}$, there exists some $1 \leq i \leq n$ such that $\inf(\pi) \subseteq F_i$ and $\inf(\pi) \cap G_i \neq \emptyset$.

- For a *Streett acceptance condition* $\mathcal{F} \subseteq 2^Q \times 2^Q$, $\pi$ is accepting if for $\mathcal{F} = \{(F_1, G_1), \ldots, (F_n, G_n)\}$ and for all $1 \leq i \leq n$, we have $\inf(\pi) \nsubseteq F_i$ or $\inf(\pi) \cap G_i = \emptyset$.

- For a *Muller acceptance condition* $\mathcal{F} \subseteq 2^Q$, $\pi$ is accepting if $\inf(\pi) \in \mathcal{F}$.

The language of $\mathcal{A}$ is defined as the set of words for which there exists a run that is accepting with respect to the type of the acceptance condition. We also call automata with a $t$-type acceptance condition $t$-automata (for $t \in$ {safety, Büchi, co-Büchi, generalised Büchi, parity, Rabin, Streett, Muller}). For Rabin automata, $|\mathcal{F}|$ is also called the *Rabin index* of the automaton. For the scope of this paper and without loss of generality, we assume that all deterministic non-safety automata have no dead-ends, i.e., for all $q \in Q$ and $x \in \Sigma$, we have $|\delta(q,x)| = 1$.

**The Rabin hierarchy** It has been proven that the set of languages representable by the following automaton types is the same (see, e.g., Grädel et al., 2002)

- deterministic Muller, non-deterministic Muller
- deterministic Streett, non-deterministic Streett
- deterministic Rabin, non-deterministic Rabin
- deterministic parity, non-deterministic parity
- non-deterministic Büchi

This set is called the $\omega$-regular languages.

Given an alphabet $\Sigma$ and some $\omega$-regular language $L \subseteq \Sigma^\omega$, there exists some number $n$ such that some deterministic Rabin automaton with $n$ acceptance pairs accepts $L$ and there does not exist a deterministic Rabin automaton with less than $n$ acceptance pairs that accepts precisely $L$. We call this number $n$ the *Rabin index* of $L$. It has been proven that the so-called *Rabin hierarchy* is strict, i.e., for every Rabin index value $n \in \mathbb{N}$, there exists some language with a Rabin index of $n$ (Kaminski, 1985).

In this paper, we pay special attention to languages with a Rabin index of 1. These strictly contain the set of languages representable by safety and deterministic Büchi or co-Büchi automata.

**Parity automata and parity games** Given a deterministic parity automaton $\mathcal{A} = (Q, \Sigma, q_0, \delta, \mathcal{F})$ with $\Sigma = 2^{(AP_I \uplus AP_O)}$, it is well-known that $\mathcal{A}$ can be converted to a parity game $\mathcal{G}$ such that $\mathcal{G}$ admits a winning strategy for player 1 (the so-called *system player*) if and only if there exists a Mealy automaton $\mathcal{M}$ reading $\Sigma_I = 2^{AP_I}$ and outputting $\Sigma_O = 2^{AP_O}$ such that the language induced by $\mathcal{M}$ is a subset of the language of $\mathcal{A}$ (see, e.g., Thomas, 2008). Furthermore, from a winning positional strategy in $\mathcal{G}$, such a Mealy automaton $\mathcal{M}$ can easily be extracted.

## 3 Obtaining deterministic automata from parts of the specification

Many specifications found in practice are of the form

$$(a_1 \wedge a_2 \wedge \ldots \wedge a_{n_a}) \to (g_1 \wedge g_2 \wedge \ldots \wedge g_{n_g}) \tag{1}$$

for some set of assumptions $a_1, \ldots, a_{n_a}$ and guarantees $g_1, \ldots, g_{n_g}$ (see, e.g., Piterman et al., 2006; Bloem et al., 2007a,b; Könighofer et al., 2009). Such a specification is typical for a case in which a single component of a bigger system is to be synthesized as some assumptions about the environment (i.e., the behaviour of the other components) can be given and the part to be synthesized in turn has to satisfy some guarantees.

Piterman et al. (2006) presented the generalised reactivity(1) synthesis approach for performing synthesis for specifications of the form stated in Formula 1. Although not explicitly stated (see, e.g., Könighofer et al., 2009), the approach can be used whenever all assumptions and guarantees are representable and given as deterministic Büchi automata. The question of how to obtain these automata from guarantees and assumptions given in some logic like LTL has been left open.

In this section, we address this problem for both the generalised reactivity(1) and the generalised Rabin(1) synthesis approaches, of which the latter will be introduced in the following section. We carefully treat the two cases and point out similarities and differences in the process of obtaining deterministic automata for the two approaches.

In the following, we abbreviate the terms "generalised reactivity(1)" by GR(1) and "generalised Rabin(1)" by GRabin(1).

### 3.1 The classical construction

The classical way of obtaining a deterministic automaton $\mathcal{A}$ from an LTL formula $\psi$ is to perform the following steps:

- Convert $\psi$ to an equivalent non-deterministic Büchi automaton $\mathcal{A}'$ (Entry points to the literature are Vardi, 1996 and Gastin and Oddoux, 2001).

- Convert $\mathcal{A}'$ to a deterministic Rabin or parity automaton using (something similar to) Safra's construction (Henzinger and Piterman, 2006; Safra, 1989).

As a result, we obtain automata with possibly high Rabin indices. For generalised reactivity(1) synthesis, we need to convert them to deterministic Büchi automata afterwards. For generalised Rabin(1) synthesis, a conversion to deterministic Rabin automata with a single acceptance pair is necessary. Furthermore, whenever this is not possible, the specification has be to discarded for being usable for GR(1)/GRabin(1) synthesis, respectively. So in both cases, additional steps have to be performed.

For the generalised reactivity(1) case, this is simple. As deterministic Rabin automata are *Büchi-type* (Kupferman et al., 2006), they can easily be converted to Büchi automata whenever possible in polynomial time (Krishnan et al., 1994). Addutionally, these Büchi automata can then be minimised (Ehlers, 2010).

For the generalised Rabin case, we can apply some algorithm for obtaining a Rabin automaton with the same language but a minimal Rabin index. Krishnan et al. (1995) describe a suitable algorithm running in time polynomial in the source automaton size.

### 3.2 Using a general LTL synthesis procedure

An alternative method for obtaining deterministic Büchi or one-pair Rabin automata equivalent to a given LTL formula $\psi$ has been given by Kupferman and Vardi (2005).

Let $\psi$ range over a set of variables $I$. The problem of obtaining an equivalent deterministic Büchi automaton can be solved by reduction to the finite-state system synthesis problem of the specification $\phi = \psi \leftrightarrow (GFout)$ with the input variable set $I$ and the output variable set $\{out\}$. Any finite-state machine satisfying the specification can be converted to a suitable Büchi automaton by duplicating all states, making one copy of each state accepting, and routing the transitions to the respective accepting states if and only if *out* is set to true.

Equivalently, obtaining a one-pair Rabin automaton from a specification $\psi$ over some variable set $I$ can be reduced to finite-state system synthesis with the specification $\phi = \psi \leftrightarrow (FGout_1 \wedge GFout_2)$, the input variable set $I$, and the output variable set $\{out_1, out_2\}$.

For performing the synthesis step in practice, any of the known algorithms can be used (see, e.g., Kupferman and Vardi, 1999, 2005; Schewe and Finkbeiner, 2007; Henzinger and Piterman, 2006; Vardi, 1996).

## 4 Performing generalised Rabin(1) synthesis by reduction to parity games

In this section, we present the core construction of the generalised Rabin(1) synthesis approach, i.e., how we transform a specification of the form

$$\psi = (a_1 \wedge a_2 \wedge \ldots \wedge a_{n_a}) \to (g_1 \wedge g_2 \wedge \ldots \wedge g_{n_g})$$

for some set of assumptions $a_1, \ldots, a_{n_a}$ and some set of guarantees $g_1, \ldots, g_{n_g}$ given in form of deterministic one-pair Rabin automata to a deterministic parity automaton with at most 5 colours that accepts precisely the words that satisfy $\psi$. The number of states of the generated automaton is polynomial in the product of the state numbers of the individual Rabin automata $a_1, \ldots, a_{n_a}, g_1, \ldots, g_{n_g}$. The generated parity automaton can then be transformed into a parity game (taking into account the partitioning of the atomic propositions into input and output bits) that is winning for player 1 if and only if there exists a Mealy automaton over the given sets of inputs and outputs such that all of its runs satisfy the specification.

Note that by the definition of Rabin acceptance, a word is accepted by a deterministic one-pair Rabin automaton $\mathcal{A} = (Q, \Sigma, q_0, \delta, \{(F, G)\})$ if and only if it is accepted by the co-Büchi automaton $\mathcal{A}_C = (Q, \Sigma, q_0, \delta, Q \setminus F)$ and the Büchi automaton $\mathcal{A}_B = (Q, \Sigma, q_0, \delta, G)$. Therefore, we can decompose a specification of the form stated above into four sets of automata:

- A set $A = \{A_1, \ldots, A_{n_1}\}$ containing the automata of the assumption conjuncts with Büchi acceptance condition
- A set $B = \{B_1, \ldots, B_{n_2}\}$ containing the automata of the assumption conjuncts with co-Büchi acceptance condition
- A set $C = \{C_1, \ldots, C_{n_3}\}$ containing the automata of the guarantee conjuncts with Büchi acceptance condition
- A set $D = \{D_1, \ldots, D_{n_4}\}$ containing the automata of the guarantee conjuncts with co-Büchi acceptance condition

For improved readability of the following description of the algorithm, by abuse of notation, we introduce $\delta$, $Q$, $q_0$, $\Sigma$, and $\mathcal{F}$ as functions mapping automata onto their components. For example, given some automaton $\mathcal{A} = (\tilde{Q}, \tilde{\Sigma}, \tilde{q}_0, \tilde{\delta}, \tilde{\mathcal{F}})$, we have $\delta(\mathcal{A}) = \tilde{\delta}$.

We furthermore assume that for all $a, a' \in A \uplus B \uplus C \uplus D$, $\Sigma(a) = \Sigma(a')$, i.e., all automata share the same alphabet.

We construct the parity automaton $\mathcal{A}' = (Q', \Sigma', \delta', q'_0, \mathcal{F}')$ as follows:

- $\Sigma'$ is chosen such that for all $a \in A \uplus B \uplus C \uplus D$: $\Sigma' = \Sigma(a)$
- $Q' = Q(A_1) \times \ldots \times Q(D_{n_4}) \times \{0, 1, \ldots, n_1\} \times \{0, 1, \ldots, n_3\} \times \mathbb{B}$
- For all $q = (q_1^A, \ldots, q_{n_4}^D, q^W, q^R, q^V) \in Q'$ and $x \in \Sigma$, we define $\delta'(q, x) = (q_1'^A, \ldots, q_{n_4}'^D, q'^W, q'^R, q'^V)$ such that:
  - For all $1 \leq i \leq n_1$: $\delta(A_i)(q_i^A, x) = q_i'^A$
  - For all $1 \leq i \leq n_2$: $\delta(B_i)(q_i^B, x) = q_i'^B$
  - For all $1 \leq i \leq n_3$: $\delta(C_i)(q_i^C, x) = q_i'^C$
  - For all $1 \leq i \leq n_4$: $\delta(D_i)(q_i^D, x) = q_i'^D$
  - $q'^W = (q^W + 1) \mod (n_1 + 1)$ if $q_{q^W}'^A \in \mathcal{F}(A_{q^W})$ or $q^W = 0$, otherwise $q'^W = q^W$.
  - $q'^R = (q^R + 1) \mod (n_3 + 1)$ if $q_{q^R}'^C \in \mathcal{F}(C_{q^R})$ or $q^R = 0$, otherwise $q'^R = q^R$.
  - $q'^V = \mathbf{true}$ if and only if (at least) one the following two conditions hold:
    * $q_W = 0$
    * for all $1 \leq i \leq n_4$, $q_i'^D \notin \mathcal{F}(D_i)$ and $q^V = \mathbf{true}$
- For all $q = (q_1^A, \ldots, q_{n_4}^D, q^W, q^R, q^V) \in Q'$, we have that $\mathcal{F}'$ maps $q$ to the least value in $c \in \{0, 1, 2, 3, 4\}$ such that:
  - $c = 4$ if for some $1 \leq i \leq n_2$: $q_i^B \in \mathcal{F}(B_i)$
  - $c \geq 3$ if $q^V = \mathbf{true}$ and for some $1 \leq i \leq n_4$, $q_i^D \in \mathcal{F}(D_i)$
  - $c \geq 2$ if $q^R = 0$.

- $c \geq 1$ if $q^W = 0$
- $q'_0 = (q_0(A_1), \ldots, q_0(D_{n_4}), 0, 0, \textbf{false})$

## 4.1 Explanation of the construction

In this sub-section, we discuss the construction of the automaton $\mathcal{A}' = (Q', \Sigma', \delta', q'_0, \mathcal{F}')$ as described above and give a correctness proof.

The states $q = (q_1^A, \ldots, q_{n_4}^D, q^W, q^R, q^V) \in Q'$ in the automaton have some components $q_1^A, \ldots, q_{n_4}^D$ that basically represent the automata of $A \uplus B \uplus C \uplus D$ running in parallel. The remaining part of the state tuples corresponds to some additional *control structure* for checking if the specification $(a_1 \wedge a_2 \wedge \ldots \wedge a_{n_a}) \rightarrow (g_1 \wedge g_2 \wedge \ldots \wedge g_{n_g})$ is satisfied. Note that adding the control structure only results in a polynomial blow-up. The parts of the control structure have the following purposes:

- The counter $q^W$ keeps track of the Büchi assumption for which an accepting state is to be visited next. The construction of this part of the parity game is essentially the same as for de-generalising generalised Büchi automata (see, e.g., Thomas, 1994, Lemma 1.2).

- The counter $q^R$ does the same for the guarantees.

- The bit $q^V$ tracks if recently accepting states for all automata in $A$ have been visited.

These counters and bits suffice for assigning colours to the states in $\mathcal{A}'$ such that the highest number occurring infinitely often along a run is even if and only if the corresponding word satisfies $(a_1 \wedge a_2 \wedge \ldots \wedge a_{n_a}) \rightarrow (g_1 \wedge g_2 \wedge \ldots \wedge g_{n_g})$. Understanding the idea behind the construction is most simple by considering the five reasons for rejecting/accepting a word individually:

1. A word should be accepted by $\mathcal{A}'$ if it is not accepted by some automaton $b \in B$ (violation of a co-Büchi assumption).

2. A word should be accepted by $\mathcal{A}'$ if it is not accepted by some automaton $a \in A$ (violation of a Büchi assumption).

3. If the assumptions are satisfied, a word should be rejected if it is not accepted by some automaton $d \in D$ (violation of a co-Büchi guarantee)

4. If the assumptions are satisfied, a word should be rejected if it is not accepted by some automaton $c \in C$ (violation of a Büchi guarantee)

5. In the remaining cases (i.e., all the assumptions and guarantees are satisfied), a word should be accepted.

It is clear from the definition of the specification that an automaton satisfying these constraints is suitable for the synthesis task. The automaton $\mathcal{A}'$ fulfils these criteria, as the following lines of thought show:

1. Assume that some automaton $b \in B$ does not accept the input/output word. In this case, rejecting states of $b$ are visited infinitely often, resulting in the colour 4 occurring infinitely often along the run. As this is the highest possible colour, the word is accepted.

2. Assume that some automaton $a \in A$ does not accept the input/output word. Without loss of generality, we can assume that all automata in $B$ accept the word, as otherwise the previous item already covers this case. So, colour 4 is not visited infinitely often.

   Since some automaton $a \in A$ does not accept the input/output word, the counter $q^W$ stalls at some point as it cycles through all automata of $A$, waiting for visits to their respective accepting states. Consequently, the value of $q^V$ can only be set from **false** to **true** finitely often. As every occurrence of colour 3 resets $q^V$ to **false** after $q^W$ has stalled and requires $q^V$ to be equal to **true** beforehand, states with colour 3 can only be visited finitely often.

Finally, colour 1 cannot be visited infinitely often as the counter $q^W$ eventually stalls.

Thus, only the colours 2 or 0 can be visited infinitely often, leading to acceptance of the word.

3. Assume that the assumptions are satisfied, but some co-Büchi-automaton $d \in D$ of the guarantee part of the specification does not accept the input word.

   In this case, as the Büchi assumptions are fulfilled, $q^W$ is set to 0 infinitely often and thus $q^V$ will be equal to **true** infinitely often. As for some $c \in C$, its rejecting state is visited infinitely often, and $q^V$ stays equal to **true** until a state with colour 3 has been visited, this implies that colour 3 occurs infinitely often. As colour 4 does not occur infinitely often (the co-Büchi assumptions are fulfilled), the input/output word is rejected.

4. Assume that the assumptions are satisfied, but some Büchi-automaton $c \in C$ of the guarantee part of the specification does not accept the input word.

   In this case, at some point during the run, the $q^R$-part of the states occurring stalls at a number $\neq 0$, i.e., the counter will not be increased or reset any longer, leading to only finitely many visits to colour 2. Since the co-Büchi assumptions and guarantees are satisfied, states with the colour 4 are only visited finitely often (see above). We can also assume that the co-Büchi guarantees are fulfilled as otherwise the previous item covers this case, so states with colour 3 are visited only finitely often.

   Thus, as the Büchi assumptions hold, the counter $q^W$ is reset infinitely often and colour 1 is the highest one occurring infinitely often, the word is rejected.

5. Assume that all guarantees and assumptions are satisfied. In this case, from some point onwards, colour 3 and 4 are never visited (as the co-Büchi assumptions and guarantees are fulfilled). The counter $q^R$ is however reset to 0 infinitely often (as the Büchi guarantees are fulfilled), which leads to infinitely many occurrences of colour 2, resulting in acceptance.

By taking these facts together, we obtain the following result:

**Theorem 1.** *The parity automaton given above accepts precisely the words $w \in \Sigma^\omega$ that satisfy the overall specification, i.e., either there exists some automaton in $A \uplus B$ that rejects $w$ or all automata in $C \uplus D$ accept $w$.*

## 5 On extending the approach to generalised Rabin($k$)-specifications with $k > 1$

The construction given in the previous section does only work for specifications with assumptions and guarantees having Rabin indices of one. A natural question to ask is: Does a similar construction also exist for guarantees and assumptions whose Rabin indices are greater than one?

In this section, we show that this is not the case. In particular, we prove the following theorem:

**Theorem 2.** *For all $k > 1$ and $c \in \mathbb{N}$, the following holds: In polynomial time, it is not possible to compute a control structure of size polynomial in $n_a + n_g$ for reducing the synthesis problem for specifications of the form*

$$(a_1 \wedge a_2 \wedge \ldots \wedge a_{n_a}) \to (g_1 \wedge g_2 \wedge \ldots \wedge g_{n_g})$$

with all assumptions $a_1, a_2, \ldots, a_{n_a}$ and guarantees $g_1, g_2, \ldots, g_{n_g}$ given as Rabin automata of index at most $k$ to the non-emptiness problem of a parity automaton with $c$ colours such that its transition structure is the parallel composition of the transition structures of the Rabin automata and the control structure (unless P=NP).

Thus, the approach presented in this paper is in some sense as far as we can get without losing its good properties. These are:

- the fact that the transition structure of $\mathcal{A}'$ is the parallel composition of the transition structures of the automata for the individual assumptions and guarantees and some control structure – this allows the efficient representation of the transition function in a symbolic way (e.g., by using binary decision diagrams, see, e.g., Drechsler and Sieling, 2001);
- the constant numbers of colours.

In the remainder of this section, we show why Theorem 2 holds. For this, we use a theorem proven by Chatterjee et al. (2007). Let $(\otimes, k, [n])$ represent the set of *generalised parity games* with an acceptance condition type $\otimes \in \{\vee, \wedge\}$ and a number $k \in \mathbb{N}$ of colouring functions, with each colouring function having a co-domain of $\{0, \ldots, n\}$. Likewise, $[n]_+$ represents colouring functions having a co-domain of $\{1, \ldots, n\}$. A play in a generalised parity game with $\otimes = \vee / \otimes = \wedge$ is accepting for player zero if for any/all of the colouring functions, the highest colour occurring infinitely often is even, respectively.

**Theorem 3** (Chatterjee et al., 2007, pp. 159). *Given a game graph $\mathcal{G}$, for objectives $\Psi$ in $(\vee, k, [3]_+)$ and $\Phi$ in $(\wedge, k, [2])$, and a vertex $v$ in $\mathcal{G}$:*

- *checking whether $v$ is a vertex winning for player 1 for $\Psi$ is NP-hard;*
- *checking whether $v$ is a vertex winning for player 0 for $\Phi$ is co-NP-hard.*

We are now ready to prove Theorem 2.

*Proof.* Assume that Theorem 2 does not hold and that we have a specification of the form $g_1 \wedge \ldots \wedge g_{n_g}$ such that all Rabin automata for $g_1, \ldots, g_{n_g}$ have the same transition structure. Since we assume that the parity automaton is the parallel composition of the transition structures of $g_1, \ldots, g_{n_g}$ and some polynomial control structure, we obtain some parity automaton with a size polynomial in $n_g$ and the number of states in the automaton of $g_1$ with a constant number of colours. Emptiness of such an automaton can consequently be decided in time polynomial in the size of the automaton of $g_1$.

This is however a contradiction to Theorem 3. To see this, note that Rabin automata with index 1 are essentially parity automata with a parity function with co-domain $\{1, 2, 3\}$. Likewise, a Streett automaton with a single acceptance pair is essentially a parity automaton with a parity function with co-domain $\{0, 1, 2\}$. All such Streett automata have a Rabin index of at most 2. Assume that we have $n_g$ Streett automata with single acceptance pairs given as specification. If they all share the same transition function, we only have to consider it once in the combined parity game. This essentially leads to a game of size polynomial in $n_g$ and the size of the transition structure of the Streett automata. Since solving this game can be done in polynomial time and the result is always a correct answer to the problem posed in Theorem 3, this would imply co-NP=P as well as NP=P. □

So, provided that NP≠P, the only way to have a similar construction with a constant number of colours would be to have an approach that does not allow the technical trick to join equivalent transition structures of the individual automata, which would be a strong indicator for unsuitability for symbolic implementations, essentially ruling out its usage for synthesis.

# 6  On application domains for the techniques described here

From a theoretical perspective, generalised Rabin(1) synthesis is a strict generalisation of generalised reactivity(1) synthesis and extends its scope by allowing co-Büchi assumptions and guarantees.

From a practical perspective, the question if the added expressivity in comparison to the approach by Piterman et al. (2006) is of practical value is natural to ask. Indeed, the benefit of the added possibility to work with co-Büchi guarantees and assumptions is not obvious. To shed light on this issue, we mention two possible application areas here:

- During the initialisation phase of a larger system implemented in hardware, the status of the system can be partly unspecified. In such a case, some components of such a system can deviate from their regular behaviour. Co-Büchi assumptions can be used to model the fact that at some point in time, such an initialisation phase is over. Additionally, co-Büchi guarantees can be used to allow deviations in the behaviour of a component of a larger system to be synthesized for a limited period of time (i.e., during the component's own initialisation phase).

- Bloem et al. (2009) discussed the benefit of adding robustness criteria to the synthesis process. In this setting, a process to be synthesized is expected to degrade gracefully on the violation of the assumptions used during synthesis. For example, consider a two-process mutex that is required to grant all requests in the same computation cycle. Formally, such a system has inputs $AP_I = \{r_1, r_2\}$ and outputs $AP_O = \{g_1, g_2\}$. Consider the specification $G(\neg r_1 \vee \neg r_2) \to G(r_1 \to g_1 \wedge r_2 \to g_2)$. It only constrains the behaviour of the system if the two processes never request a grant at the same time. In case of a violation of this constraint, however, no restriction on the behaviour of the mutex is made. Bloem et al. (2009) argue that in practice, most systems are somewhat robust against such assumption violations. For example, the component to be synthesized could continue to respond to requests in the correct way after a violation of the assumption, i.e., whenever only one request is given at the same time, the respective grant is given.

  For the *qualitative* version of robust synthesis, co-Büchi specifications are a natural way to express such degradable parts of the assumption or guarantee. For example, a finite-state system satisfying $FG(\neg r_1 \vee \neg r_2) \to FG(r_1 \to g_1 \wedge r_2 \to g_2) \wedge G(\neg g_1 \wedge \neg g_2)$ can only violate the responsiveness guarantee infinitely often if the assumption $\neg r_1 \vee \neg r_2$ is violated infinitely often. Since it only has a finite number of states, it is thus forced to return to normal behaviour after a limited amount of time after some computation cycle in which $\neg r_1 \vee \neg r_2$ is violated, which makes it a valid solution.

  Bloem et al. (2009) presented algorithmic solutions for the robust synthesis problem for safety specifications. They leave an extension of their techniques to the liveness case as an open problem. As with generalised Rabin(1) synthesis, we are able to handle such specifications, the technique presented here is a suitable solution to this open problem.

# 7  Conclusion

In this paper, we have presented generalised Rabin(1) synthesis as a strict generalisation of generalised reactivity(1) synthesis and showed that it shares its good algorithmic properties. This increases the practical applicability of the approach and is thus a big step forwards towards synthesis from large specifications. We also showed that the concept cannot be extended further without losing its good algorithmic properties.